\documentclass[prl,aps,twocolumn,floatfix,amsmath,amssymb,superscriptaddress]{revtex4-1}
\usepackage{latexsym}
\usepackage{graphicx}
\usepackage{tabularx}
\usepackage{times}
\usepackage[usenames,dvipsnames]{color}
\usepackage{amsmath}
\usepackage{dcolumn}
\usepackage{latexsym,amsmath,amssymb,bm,euscript}
\usepackage[
	colorlinks=true,
	urlcolor=BlueViolet,
 citecolor=Maroon,
	plainpages=false,
 	pdfpagelabels,
 	bookmarksnumbered
 	]{hyperref}

\renewcommand{\i}{\mathrm{i}}

\renewcommand{\(}{\left(}
\renewcommand{\)}{\right)}
\def\beq{\begin{equation}}
\def\eeq{\end{equation}}

\begin{document}\renewcommand{\i}{\mathrm{i}}

\title{Anomalous Quasiparticle Symmetries and Non-Abelian Defects on Symmetrically Gapped Surfaces of Weak Topological Insulators}

\author{David F. Mross}
\affiliation{Department of Physics and Institute for Quantum Information and Matter, California Institute of Technology, Pasadena, CA 91125, USA}
\author{Andrew Essin}
\affiliation{Department of Physics and Institute for Quantum Information and Matter, California Institute of Technology, Pasadena, CA 91125, USA}
\author{Jason Alicea}
\affiliation{Department of Physics and Institute for Quantum Information and Matter, California Institute of Technology, Pasadena, CA 91125, USA}
\affiliation{Walter Burke Institute for Theoretical Physics, California Institute of Technology, Pasadena, CA 91125, USA}
\author{Ady Stern}
\affiliation{Department of Condensed Matter Physics, Weizmann Institute of Science, Rehovot, 76100, Israel}

\begin{abstract}
We show that boundaries of 3D weak topological insulators can become gapped by strong interactions while preserving all symmetries, leading to Abelian surface topological order.  The anomalous nature of the weak topological insulators manifests itself in a non-trivial action of symmetries on the quasiparticles; most strikingly,  translations change the anyon types in a manner impossible in strictly 2D systems with the same symmetry.  As a further consequence, screw dislocations form non-Abelian defects that trap $\mathbb{Z}_4$ parafermion zero modes.
\end{abstract}
\maketitle

\section{Introduction}
Electronic topological insulators \cite{FuTI,MooreTI,RoyTI,KaneReview,QiReview} display numerous exotic properties already at the single-particle level, most famously protected surface metallicity.  Much of the richness in these systems emerges from the interplay between symmetry and topology.  Recently interactions among surface electrons have been found to further enlarge the possibilities.  In a \emph{strong} topological insulator (STI) the surface spectrum for weakly interacting electrons obeying time reversal ${\cal T}$ and charge conservation symmetry features a single Dirac cone. Remarkably, strong interactions can fully gap the STI surface without violating symmetries \cite{BondersonTO,WangTO,ChenTO,MetlitskiTO} (as anticipated earlier \cite{levinburnellkochstern}). The symmetrically gapped phases realize non-Abelian topological order---the simplest incarnation being the `T-Pfaffian'---and can be viewed as descending from novel gapless states \cite{STO,MetlitskiVishwanath2015,WangSenthil2015}.
Similar conclusions hold for bosonic topological insulators \cite{AshvinSenthil}, topological superconductors \cite{MetlitskiTO2}, and topological crystalline insulators \cite{QiFu2015}.  (Not all topological systems, however, admit a symmetric gapped boundary \cite{WangSenthil2014}.)

\begin{figure}[ht]
\includegraphics[width=\columnwidth]{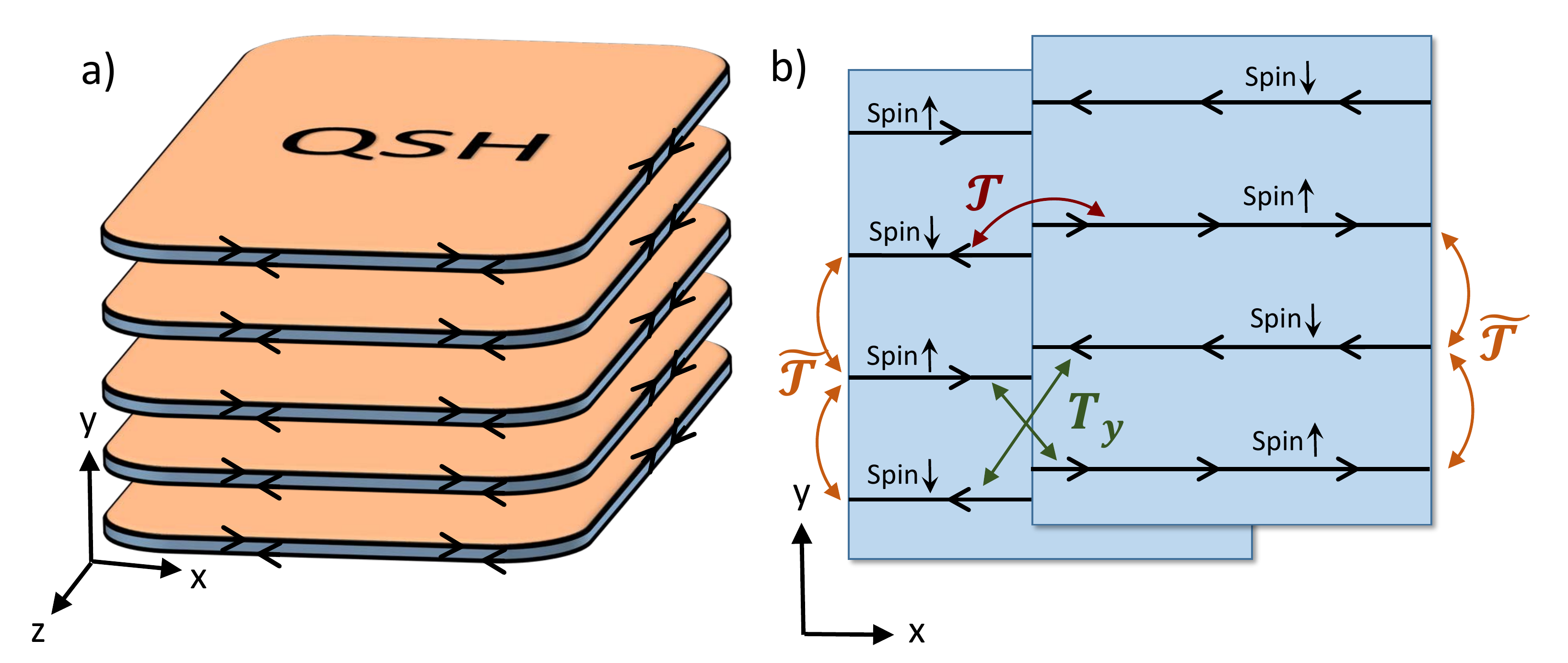}
\caption{(color online) (a) Weak topological insulator built from quantum spin Hall layers.  Interlayer tunneling yields two symmetry-protected surface Dirac points at momenta $(q_x,q_y) = (0,0)$ and $(0,\pi)$.  (b) Weak topological insulator surface viewed as a bilayer.  }
\label{fig:stack}
\end{figure}

We explore for the first time the fate of strongly correlated \emph{weak} topological insulator (WTI) surfaces.  These systems comprise ideal settings where one can controllably explore the influence of additional symmetries, which enrich the surface topological order that we identify in subtle ways and yield an interesting interplay with lattice defects.  A WTI may conveniently be decomposed into a stack of quantum spin Hall (QSH) insulators \cite{FuTI,MooreTI,RoyTI} with electrons from the helical edges tunneling between layers; see Fig.~\ref{fig:stack}.  Provided the system preserves time reversal $\mathcal{T}$, charge conservation, and layer translation symmetry $T_y$, the \emph{non-interacting} WTI surface hosts two massless Dirac cones at distinct momenta \footnote{The WTI surface exhibits low-energy properties with surprising resilience to spatial disorder \cite{RingelKrausStern2012,MongBardarsonMoore2012,FuKane2012}, which is apparently related to the even-odd effect discussed below.  We nevertheless assume translation invariance except for isolated defects.}.

\begin{figure}[ht]
\includegraphics[width=\columnwidth]{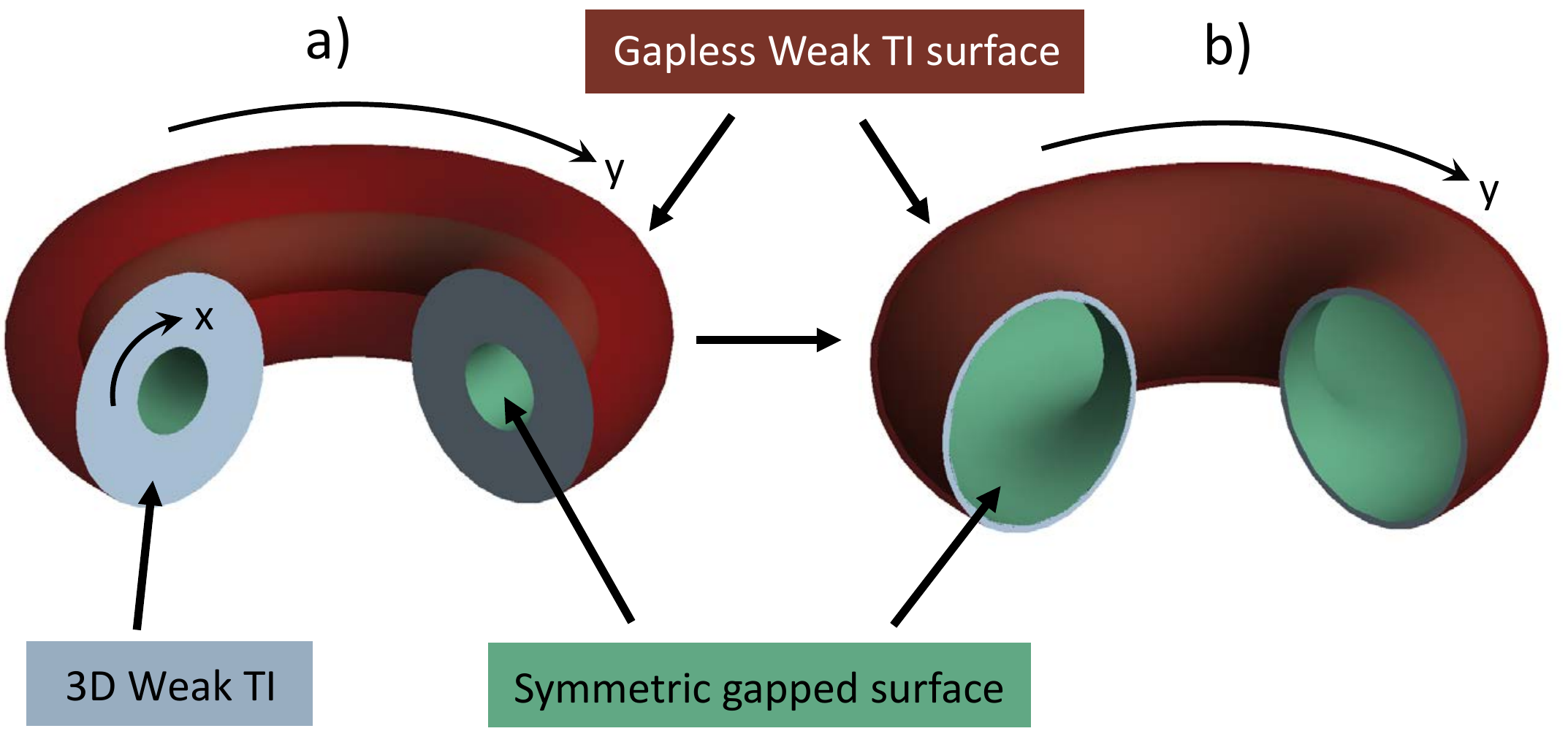}
\caption{(color online) (a) Thickened torus of weak topological insulator with symmetric topologically ordered interior and gapless exterior.  (b) Two-dimensional limit where the thickness shrinks to zero.  The topological order must be anomalous; otherwise one is left with an `impossible' 2D band structure.}
\label{fig:tori}
\end{figure}

{Four considerations are useful for anticipating the topological order that emerges when interactions gap the WTI surface without violating these symmetries. 
First, on very general grounds the topological order must be anomalous, i.e., forbidden in strictly 2D isosymmetric systems.  To see this consider the  thickened torus of WTI depicted in Fig.~\ref{fig:tori}(a), and gap the interior surface by interactions but leave the exterior gapless.  Upon shrinking the torus's thickness a strictly 2D system emerges as in Fig.~\ref{fig:tori}(b).  If the gapped surface was non-anomalous, one could simply strip away the topological order, leaving a symmetric 2D system with an `impossible' band structure \footnote{Two Dirac cones occurring at different momenta cannot appear in spinful, time-reversal- and translation-symmetric 2D systems} ---a contradiction.

The second consideration regards a domain wall separating the topologically ordered state from a ferromagnetically gapped surface region. The magnetized region carries a non-zero thermal Hall conductivity and thus the domain wall must host gapless modes. In the STI case the thermal Hall conductivity would be half-integer (in units of $\pi^2 k_B^2 T/3h$); the gapless mode's central charge must also then be half-integer, necessitating a non-Abelian topological order.  By contrast, the two Dirac cones present for the WTI imply an integer central charge, suggesting an Abelian minimal topological order.}

{The third consideration seemingly contradicts the second by raising expectations for a non-Abelian surface state:  Since interactions can drive the STI's \emph{single} Dirac cone into the non-Abelian T-Pfaffian \cite{BondersonTO,WangTO,ChenTO,MetlitskiTO}, the WTI's two surface Dirac cones may naturally be gapped independently in the same manner.  The resulting non-Abelian surface hosts a large number of quasiparticles---two copies of each T-Pfaffian excitation.  As we confirm below, however, the quasiparticle content can be reduced via a symmetry-preserving `anyon condensation' transition that confines the non-Abelian excitations leaving only an Abelian set, consistent with the argument above.  

The fourth consideration results from viewing the WTI as a stack of QSH insulators.  Any finite stack may be viewed as two dimensional, with an even-odd effect: the system forms a 2D topological insulator with an odd number of layers but a trivial 2D insulator otherwise.  Since the 2D topological insulator edge cannot be gapped without breaking $\mathcal{T}$ or charge conservation, this even-odd effect should also appear when interactions gap the stack's surface to form topological order in the limit of infinitely many layers. }

With these points in mind let us investigate gapping of the WTI in detail.  First we expound on the relation to the STI by equivalently describing the WTI surface as a bilayer system, partitioning the right- and left-movers from each QSH edge as in Fig.~\ref{fig:stack}(b).  
Here time-reversal ${\cal T}$ and translations $T_y$ interchange the two layers while $\tilde{\cal T}={\cal T}T_y$ does not.  
Each layer represents the surface of an antiferromagnetic topological insulator \cite{AFTImong,AFTIfang}---which also supports a single Dirac cone---and maps to the setup considered in Ref.~\cite{STO} for studying the correlated STI surface.  Interactions can drive each layer into an electrically insulating, symmetry-preserving `composite Dirac liquid' that hosts a single Dirac cone built from emergent \emph{neutral} fermions that carry a fictitious `pseudocharge' $\tilde e$ (which is not microscopically conserved).  Intra-cone Cooper pairing of neutral fermions in the bilayer generates `double T-Pfaffian' non-Abelian topological order.  One can conveniently view the resulting anyons as defects in the paired condensates; in particular, an $hc/2\tilde e$ vortex carries \emph{physical} charge $e/4$ and binds a Majorana zero mode.  The so-far decoupled layers host independent $hc/2\tilde e$ vortices.  Our analysis below shows how to obtain an Abelian state for which individual $h/2\tilde e$ vortices are confined, yielding $e/2$ as the minimal charge.
The Abelian state further includes a neutral quasiparticle---an interlayer $\pm e/4$ dipole---that acquires a $\pi/2$ phase when encircling an $e/2$ excitation.

\begin{figure}
\includegraphics[width=\columnwidth]{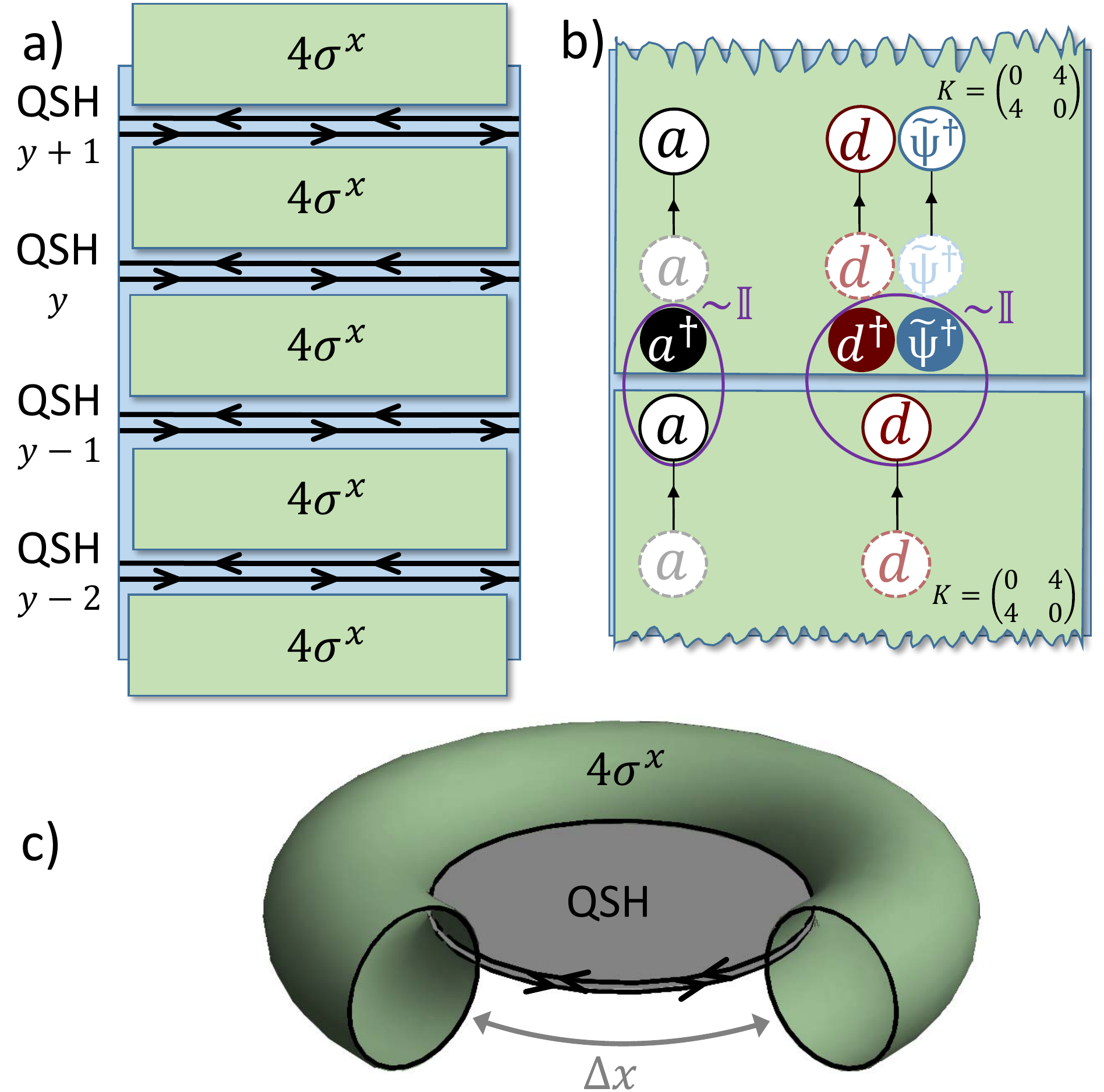}
\caption{(color online) (a) Weak topological insulator surface dressed with 2D topologically ordered `plates'.  (b) Dragging an anyon across plates leaves behind `invisible' operators at the interface that get absorbed into a condensate.  The condensates allow the $e/2$ $a$ quasiparticle to pass freely between plates while the neutral $d$ quasiparticle acquires a neutral fermion and thus changes anyon type. (c) Setup for discussing weak symmetry breaking. }
\label{fig:weaksb}
\end{figure}

We now put this discussion on firmer footing.  To facilitate gapping the WTI we imagine patterning the surface with 2D topologically ordered `plates' as in Ref.~\cite{STO}.  Figure~\ref{fig:weaksb}(a) depicts the decorated structure.   Each interface (labeled $y$) contains a helical QSH mode and two sets of gapless edge states from the adjacent plates, one from above and one from below.  
We wish to judiciously select the plates such that $(i)$ local interactions within a given interface can remove all gapless modes without breaking any symmetries and $(ii)$ the surface topological order with minimal degeneracy on a torus appears.  Note that time-reversal symmetry constrains the latter degeneracy to be the square of an integer \cite{LevinStern}.

The interfaces to be gapped are described by a K-matrix $K$ and charge vector $Q$, which specify the statistics and charges of low-energy fields, along with a vector $X$ that distinguishes Kramers singlets from doublets (for details see \cite{LevinStern}).  More precisely, we have
\begin{equation}
K=
\begin{pmatrix} K_{h} & 0 & 0 \\ 0 & K_p & 0 \\ 0 & 0 & -K_p\end{pmatrix},~~~Q = \begin{pmatrix} q_h \\ q_p \\ q_p\end{pmatrix},~~~X = \begin{pmatrix} \chi_h \\ \chi_p \\ \chi_p\end{pmatrix}
\label{bigK}
\end{equation}
where the `$h$' and `$p$' subscripts indicate quantities for the helical QSH modes and plates, respectively.  For the QSH sector $K_{h} = \sigma^z$ (here and below $\sigma^a$ denote Pauli matrices), $q_{h}=(1,1)$, and $\chi_{h}=(0,1)$.  
For the plates, time reversal demands an even-dimensional $K_{p}$.  We assume the smallest two-dimensional $K_p$, which can be either fermionic or bosonic.  We focus on the latter since we find that the fermionic case does not permit time-reversal-invariant gapping of the interface.  The bosonic case allows two distinct possibilities: $(i)$~$K_p=m\sigma^x$, $q_p=(0,2)$, $\chi_p=(r,0)$ or $(ii)$~$K_p = m \sigma^z$, $q_p=(2,2)$, $\chi_p=(r,0)$ with $m$ an even integer and $r = 0$ or 1. Either possibility yields a minimal charge excitation of $e^*=2/m$. By the criterion of Ref.~\cite{LevinStern} the interface may be symmetrically gapped when $\frac{1}{e^*}\chi^TK ^{-1}Q$ is even. It follows that the smallest possible value of $m$ is four, and that the value of $r$ does not affect the interface's gappability.  Hereafter we set $r = 0$ for concreteness and focus on $K_p = 4\sigma^x$, which relates to the `double T-Pfaffian' discussed above (see Appendix).

To specify the gap-opening interactions we introduce low-energy fields describing a given interface $y$.  Right/left-moving QSH electron operators are $\psi_{R/L,y} \equiv e^{i \varphi_{R/L,y}}$.  We use subscripts $+$ and $-$ to denote fields from the adjacent upper and lower plates.  Operators $a_{\pm,y}\equiv e^{i\phi_{a\pm,y}}$ and $d_{\pm,y} \equiv e^{i\phi_{d\pm,y}}$ then respectively create self-bosonic charge-$e/2$ and neutral excitations with time-reversal properties $a_{\pm,y}\rightarrow a_{\pm,y}$ and $d_{\pm,y}\rightarrow d_{\pm,y}^\dagger$.  These quasiparticles exhibit mutual statistics $e^{i \pi/2}$, implying that $a_{\pm,y}^4$ and $d_{\pm,y}^4$ represent local bosons.  Interactions 
\begin{align}
&\(\psi_{R}\psi_{L}\)^2 \(a_- a_+\)^4+H.c. \sim \cos 4 \theta_c\label{eqn.cgap}\\
&\(\psi_{R}^\dagger \psi_{L}\)^2 \(d_-^\dagger d_+\)^4+H.c. \sim \cos 4 \theta_s\label{eqn.sgap}\\
&\(a_-^\dagger a_+\)^4+H.c. \sim \cos 4 \theta_n\label{eqn.ngap},
\end{align}
are therefore physical.  (We suppress $y$-dependence whenever unneeded.)  The fields $\theta_{c,s,n}$ defined above obey trivial commutation relations and can therefore be simultaneously pinned to gap the interfaces.  Moreover, the interactions preserve both $\mathcal{T}$ and charge conservation. Thus uniformly condensing $\langle e^{i \theta_{c,s,n}}\rangle \neq 0$ respects all symmetries.

Determining the resulting surface topological order requires identifying the deconfined anyons.  The plates carry 16 quasiparticles built from combinations of $a$ and $d$.  What happens when we drag a quasiparticle from one plate to another as in Fig.~\ref{fig:weaksb}(b)?  Consider first dragging an $a$ charge-$e/2$ anyon.  Since fractional excitations cannot directly cross between plates, this process leaves a dipole described by $a_{y_-}a_{y_+}^\dagger \sim e^{i \theta_n}$ at the interface as the figure illustrates.  However, the condensate $\langle e^{i\theta_n}\rangle$ readily absorbs the dipole---which is effectively invisible---negating any energy cost.  The $a$ quasiparticle thus propagates freely across the surface. 

Similarly dragging a neutral $d$ anyon between plates does not simply leave behind a $d_{y_-}d_{y_+}^\dagger$ dipole since this object is uncondensed.  To specify its fate we define a neutral fermion
 \begin{align}
 \tilde \psi_{R/L} = \psi_{R/L} a^2.
 \label{eqn.neutraldef}
 \end{align}
The condensates identify $\tilde \psi_R$ and $\tilde \psi_L^\dagger$; we therefore refer to both as simply $\tilde \psi$.
When $d$ crosses an interface it leaves the combination $d_{y_-}d_{y_+}^\dagger\tilde \psi$---which is condensed---and turns into a \emph{different} anyon corresponding to $d$ augmented by the neutral fermion $\tilde \psi$.
Thus quasiparticles $\tilde d$ given by
\begin{align}
             \tilde d = \begin{cases}
               d, \ \ \ \ \ \ \ \ &\text{even plates}\\
               d \tilde \psi^\dagger, \ \ \ \ &\text{odd plates}
             \end{cases}
\end{align}
also propagate freely across the surface. 
Remarkably, translations act nontrivially on these anyons:
\begin{align}
T_y \tilde d T^{-1}_{y}=\tilde d \tilde \psi \label{eqn.transdipole}.
\end{align}
This property, which manifests an even-odd effect, crucially distinguishes the symmetrically gapped WTI surface and the topological order formed by individual plates.  Table \ref{tab.topodata} summarizes the topological data for the surface.

\begin{table}
\caption{Topological data for the fundamental anyons $\tilde d$ and $a$ in the symmetrically gapped weak topological insulator surface.}
\begin{tabular}{| c| c| c| c | c| c|}
\hline
   Anyon & Charge &$ {\cal T}$ &$ T_y$ & Braid with $\tilde d$ &Braid with $a$\\
   \hline
  $\tilde d$ & $0$ & $\tilde d^*$ & $\tilde d a^2(\times {\text{electron}})$ &0 & i \\
  $a$ & $e/2$ & $a$ & $a$ & i &0\\
  \hline
\end{tabular}
\label{tab.topodata}
\end{table}

Breaking either translation or time-reversal invariance allows the topological order to appear in strict 2D.  Absent the former, the gapped WTI surface becomes equivalent to a single plate, i.e., the topological order is faithfully described by $K_p = 4\sigma^x, q_p = (0,2), \chi_p = (0,0)$.  If instead we weakly break time reversal the surface should be described by a generally different K-matrix and charge vector (that again yield $\sigma_{xy} = 0$ due to the energy gap), as well as a translation matrix $M_y$.  The latter must obey $M_y^2 = 1$, preserve the K-matrix, and encode Eq.~\eqref{eqn.transdipole} while acting trivially on $a$ quasiparticles and electrons. Trivial action means that a translated excitation at most acquires a local boson that transforms trivially under all symmetries. For example, the operation
\begin{align}
M_y \psi_R M_y^{-1} = a^4 \psi_R^\dagger \sim \psi_R \(\psi_R^\dagger\psi_L a^4 \psi_R^\dagger \psi_L^\dagger\)\label{eqn.translate}
\end{align}
multiplies $\psi_R$ by the local boson in parenthesis. Since this boson is odd under time reversal,  Eq.~(\ref{eqn.translate}) legitimately implements translations only in the absence of time-reversal symmetry. 
One can indeed view the $\mathcal{T}$-broken surface as a 2D system described by \footnote{The time-reversed version follows by swapping the $1$ and $-1$ entries in $K$ while leaving $M_y$ fixed.}
\begin{align}
K = \begin{pmatrix} 
0 & 4 & 0 &  0\\ 
4 & 0 & 0 &  0 \\ 
0 & 0 & 1 &  0 \\ 
0 & 0 & 0 & -1\end{pmatrix}, q =\begin{pmatrix} 2 \\ 0 \\1 \\1 \end{pmatrix},
M_y = \left(
\begin{array}{cccc}
 1 & -2 &  4 & 0\\
 0 &  1 &  0 & 0\\
 0 &  1 & -1 & 0\\
 0 &  0 &  0 & 1\\
\end{array}
\right)
\label{eqn.tbroken}
\end{align}
with translations implemented in precisely this way.  
Notice that if we ignore $M_y$ Eq.~\eqref{eqn.tbroken} could equally well describe a time-reversal-invariant 2D phase.  However, enforcing translation symmetry through $M_y$ violates time reversal; see Eq.~\eqref{eqn.translate}.  In fact generalizing such 2D realizations to implement time-reversal \emph{and} translation symmetry as in the WTI surface is impossible following our earlier general arguments (Fig.~\ref{fig:tori}).

It is illuminating to discuss the gapped WTI when the system consists of a finite stack of $N$ QSH layers.  The surface is then quasi-1D and hence technically cannot sustain the required topological order.  Indeed, this case reveals
a subtlety regarding time reversal and the possibility of weak symmetry breaking \cite{weaksb}.  As a primer consider Fig.~\ref{fig:weaksb}(c) where a cylindrical plate  `wraps around' a single QSH edge, leaving a gapless helical region of length $\Delta x$.  The QSH/plate interface is identical to that considered above, and the same interactions (\ref{eqn.cgap})--(\ref{eqn.ngap}) can open a gap---ostensibly without breaking symmetries.  Furthermore the circular edge of the plate can ostensibly \emph{also} be symmetrically gapped (either by finite-size effects or interactions \cite{LevinStern}).  Interestingly, symmetry-breaking must nevertheless occur \cite{weaksb}:  A right-moving electron from the gapless QSH edge cannot penetrate into the adjacent gapped segments and must reflect into an opposite-spin left-mover, indicating spontaneous time-reversal symmetry breaking.  

Using Eq.~\eqref{eqn.sgap} one can express the magnetization at the gapless region's endpoints as $\langle \psi_L^\dagger \psi_R\rangle \sim e^{i 2 \theta_s} \langle (d_-^\dagger d_+)^2 \rangle$.
Three cases exist: $(i)$ When interactions gap the plate's circular edges the expectation value on the right side must be circumference-independent. $(ii)$ When they are gapped by finite-size effects, $\langle d_-^\dagger d_+\rangle$ and thus the magnetization decay as a power-law in the cylinder circumference $L$. $(iii)$ Finally, when the entire QSH edge is gapped ($\Delta x \rightarrow 0$) the circular edges are simply absent and $\langle  d_-^\dagger d_+ \rangle \sim e^{-L/\xi}$ with a length $\xi$ set by the plate's bulk quasiparticle gap.  This last case corresponds to the setup examined in Ref.~\cite{weaksb}.
Consider next the $N = 2$ generalization of Fig.~\ref{fig:weaksb}(c) where plates arranged into a cylinder gap two QSH layers.  The above argument for spontaneous time-reversal symmetry breaking no longer holds since a right-moving QSH electron from one layer can backscatter into the other---which $\mathcal{T}$ symmetry readily permits.  

These two examples signify an even-odd effect. For $N$ layers with periodic boundary conditions along $y$ the local magnetization at an interface analogously reads
\begin{align}
\langle \psi_{L,y}^\dagger \psi_{R,y} \rangle \sim\left\langle \(d_{-,y}^\dagger { d}_{+,y}\)^2\right\rangle.
\label{wtimagnetization}
\end{align}
A finite expectation value generically arises if a $d$ quasiparticle from just above the interface can propagate intact to the bottom side of the interface.  (Direct tunneling is disallowed, so the quasiparticle must take the `long way' around.)  The issue is subtle since $d$ acquires a neutral fermion $\tilde \psi$ when crossing an interface; recall Fig.~\ref{fig:weaksb}(b).  With even $N$ the initial $d$ ends up dressed by $\tilde \psi$ when it reaches the bottom of the interface, and the magnetization thereby vanishes.  By contrast, for odd $N$ the $d$ quasiparticle boldly arrives undressed yielding a finite value.  If the entire surface is gapped this expectation value decays exponentially with $N$, while with adjacent gapless modes [similar to Fig.~\ref{fig:weaksb}(c)] a power-law emerges.

The fact that time-reversal symmetry must be (weakly) broken in an odd-layer system follows from very general considerations and thus comprises a useful consistency check.  The surface-state spectrum for an odd number of non-interacting QSH layers is gapless with an energy splitting $\Delta E \propto 1/N$ to the next band. Thus for energies below $\Delta E$ the system maps to a single 2D QSH state. Now consider such a finite WTI and assume that interactions gap the surface only for $x>0$.  Electrons from the lowest band must backscatter with unit probability upon approaching the gapped region; since $\Delta E \propto 1/N$ the magnetization must also scale as a power-law (or slower) to ensure this outcome.  This is exactly what we found.

\begin{figure}[ht]
\includegraphics[width=.6\columnwidth]{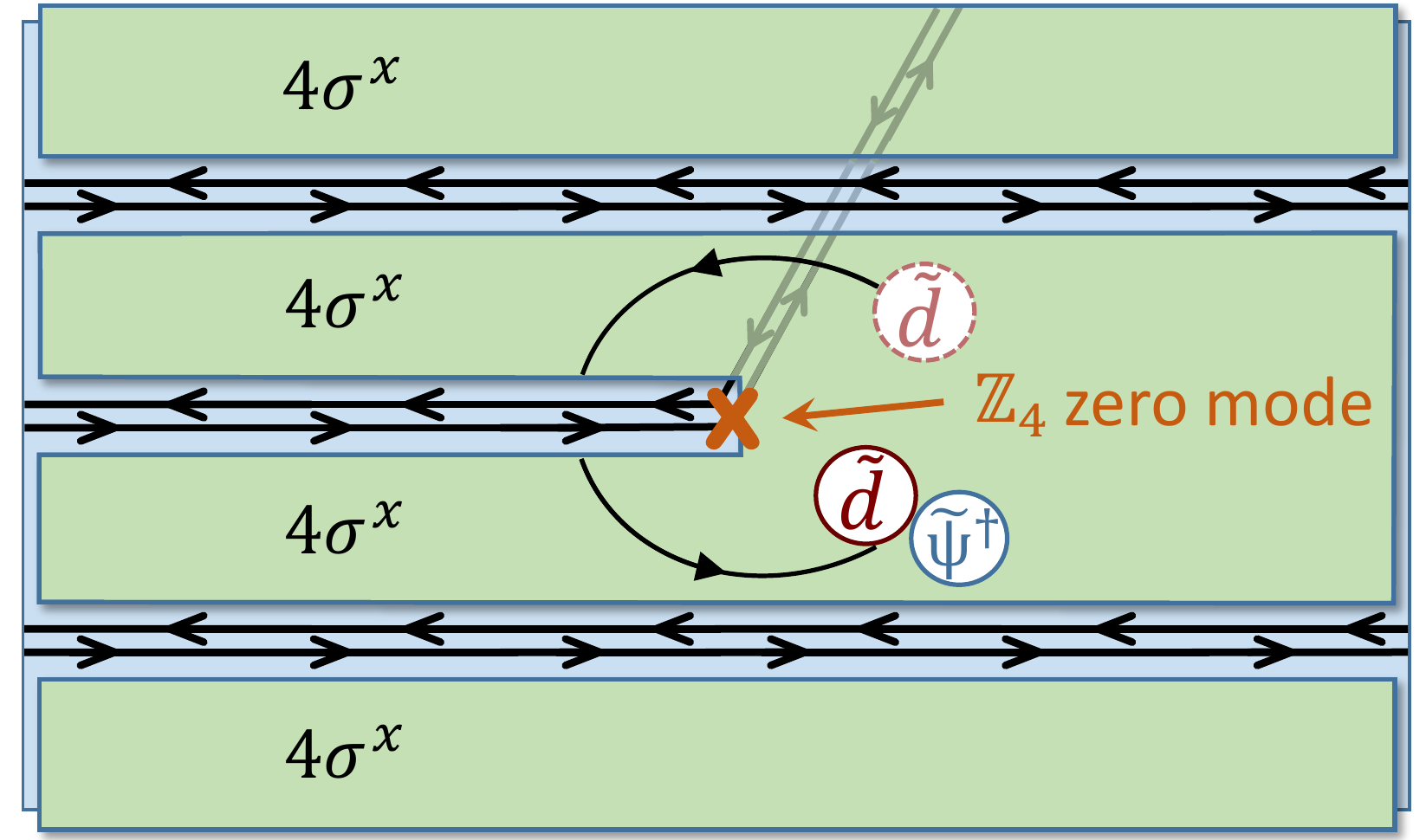}
\caption{(color online) Weak topological insulator surface with a screw dislocation. Upon encircling the dislocation $\tilde d$ anyons acquire a neutral fermion, indicating a zero mode bound to the defect. }
\label{fig:defect}
\end{figure}

The non-trivial action of translation symmetry on $\tilde d$ anyons yields interesting consequences for lattice defects.  In a WTI screw dislocations terminating at position $x_0$ on the surface (as in Fig.~\ref{fig:defect}) bind a helical QSH edge state that penetrates into the bulk \cite{ashvinscrew}.  When interactions gap the WTI boundary, electrons from the bulk helical modes must backscatter at the surface.  Such a defect thus locally violates time-reversal symmetry---yet another manifestation of weak symmetry breaking.  The impact on surface anyons is even more striking: When $\tilde d$ encircles the termination point as sketched in Fig.~\ref{fig:defect} it changes anyon type and acquires a neutral fermion.  This suggests that the dislocation forms an extrinsic non-Abelian defect that traps a nontrivial zero mode (similar effects arise in \cite{Bombin,maissamqiwormhole,TeoHughes,Benalcazar,TeoRoyChen2014b}).  To confirm this, one may formally associate the QSH edge fields $\psi_{R/L}$ that enter the bulk with the surface modes at $x>x_0$ and add a two-particle backscattering term $(\psi_R^\dagger \psi_L)^2 + H.c.$ to capture the spontaneous time-reversal symmetry breaking.  The Appendix derives the following effective Hamiltonian density that describes the defect,
\begin{align}
\mathcal{H}=\tilde \Delta\Theta(x_0-x)  \tilde \psi_R \tilde \psi_L +\tilde u\Theta(x-x_0)(\tilde \psi_R^\dagger \tilde \psi_L)^2+H.c.,
\label{eqn.zeromode}
\end{align}
with $\tilde \psi_{R/L}$ defined in Eq.~(\ref{eqn.neutraldef}).  (Note however that at $x>x_0$ we no longer have $\tilde \psi_{R}\sim\tilde \psi_{L}^\dagger$.)  The $\tilde \Delta$ and $\tilde u$ terms respectively arise from Eq.~\eqref{eqn.cgap} and the two-particle backscattering upon taking into account condensates involving the plates.  References~\cite{kanezhang,Orth} analyzed precisely Eq.~\eqref{eqn.zeromode} and showed that the defect hosts a $\mathbb{Z}_4$ parafermion zero mode.  The `$\mathbb{Z}_4$-ness' reflects the two possible values for the spontaneous magnetization and (neutral) fermion parity.

To conclude, we explored strongly interacting WTI surfaces using a quasi-1D formulation that permits full analytical control.  We found that the surface can become gapped by entering an Abelian topologically ordered state with several unusual features.  First, symmetries act on quasiparticles in a manner forbidden in purely 2D setups.  Second, an interesting even-odd effect previously known for non-interacting electrons persists in the topologically ordered surface: For a WTI composed of an odd number of QSH systems, `weak symmetry breaking' leads to a magnetization exponentially small in the number of layers. Third, lattice defects in the Abelian topologically ordered surface exhibit a non-Abelian structure, which may be viewed as a manifestation of the anomalous symmetry properties of the quasiparticles.  We expect such features to persist quite generally in weak topological phases assembled from 2D symmetry-protected topological states.  

{\bf Acknowledgments.} We gratefully acknowledge Xie Chen and Michael Levin for valuable discussions.  
This work was supported by the NSF through grant DMR-1341822 (JA); the Alfred P. Sloan Foundation (JA); the Caltech Institute for Quantum Information and Matter, an NSF Physics Frontiers Center with support of the Gordon and Betty Moore Foundation (DFM, AE, JA); the Walter Burke Institute for Theoretical Physics at Caltech; Microsoft Station Q (AS), the European Research Council under the European Union�s Seventh Framework Programme (FP7/2007-2013) / ERC Project MUNATOP (AS), the US-Israel Binational Science Foundation and the
Minerva Foundation (AS).

\bibliography{SurfaceTO}
\appendix
\section{Derivation of Defect Hamiltonian}\label{zeromode}
We now derive the Hamiltonian in Eq.~(\ref{eqn.zeromode}) that describes the dislocation depicted in Fig.~\ref{fig:defect}.  Continuing to denote the defect position by $x_0$, the gap-opening interactions for $x<x_0$ are given by
\begin{align}
&\hat {\mathcal O} _c\equiv \(\psi_{R}\psi_{L}\)^2 \(a_- a_+\)^4+H.c. \nonumber \\
&\hat {\mathcal O} _s\equiv \(\psi_{R}^\dagger \psi_{L}\)^2 \(d_-^\dagger d_+\)^4+H.c.~~~~~~~~~(x<x_0)\\
&\hat{\mathcal O} _a\equiv\(a_-^\dagger a_+\)^4+H.c. , \nonumber
\end{align}
which are precisely the terms invoked in the main text to generate symmetric topological order.  At $x > x_0$ the upper and lower plates fuse together \emph{without} the aid of an intervening QSH edge mode---which detours into the bulk at $x_0$ as the figure illustrates.  As in the main text we use $x>x_0$ to denote both the QSH edge modes that enter the bulk and plate fields at the surface.  The following perturbations describe the fusion of the plates in this region,
\begin{align}
  &\hat {\mathcal O} _d\equiv  \(d_-^\dagger d_+\)^4+H.c. \nonumber \\ 
  &\hat{\mathcal O} _a\equiv\(a_-^\dagger a_+\)^4+H.c., ~~~~~(x>x_0).
\end{align}
When relevant the above terms catalyze condensation of $\langle d_-^\dagger d_+\rangle$ and $\langle a_-^\dagger a_+\rangle$, allowing the anyons to seamlessly pass between plates as desired.  It remains to specify the fate of the QSH edge fields $\psi_{R/L}$ that bleed into the bulk.  We will add a perturbation (assumed relevant)
\begin{align}
&\hat {\mathcal O} _\psi \equiv \(\psi_{R}^\dagger \psi_{L}\)^2 + H.c.,~~~~~~(x>x_0)
\end{align}
for those modes; this catalyzes a magnetization $\langle\psi_R^\dagger \psi_L\rangle\neq 0$ without changing the symmetries of the problem since the defect spontaneously breaks time-reversal symmetry as discussed earlier.  

Notice that $\hat{\mathcal O} _a$ is present on both sides of the defect.  We can therefore set $a \equiv a_- \sim a_+$ in all remaining terms.  Furthermore, $\hat {\mathcal O} _s$ at $x<x_0$ is simply a product of terms present in $\hat {\mathcal O}_d$ and $\hat {\mathcal O}_\psi$ for $x>x_0$.  Minimizing the energy therefore requires fixing $\langle \psi_{R}^\dagger \psi_{L}(d_-^\dagger d_+)^2\rangle$ to a uniform constant everywhere.  (Otherwise there will be an energy cost when the expectation value `twists' at $x_0$.)  It follows that $d_-^\dagger d_+$ is slaved to the magnetization $\psi_R^\dagger \psi_L$ in the region $x>x_0$.  

All the interesting defect physics has now been distilled into the perturbations $\hat {\mathcal O} _c$ and $\hat {\mathcal O}_\psi$.  In terms of neutral fermions $\tilde\psi_{R/L} \equiv \psi_{R/L} a^2$ we have
\begin{equation}
  \hat {\mathcal O} _c \sim \left(\tilde\psi_R \tilde \psi_L \right)^2 + H.c.
\end{equation}
and
\begin{equation}
  \hat {\mathcal O} _\psi \sim \left(\tilde\psi_R^\dagger \tilde \psi_L \right)^2 + H.c.
\end{equation}
Focusing on these crucial pieces, the defect Hamiltonian density becomes
\begin{align}
& \mathcal{H}' =\tilde \Delta\Theta(x_0-x)  (\tilde \psi_R \tilde \psi_L)^2 +\tilde u\Theta(x-x_0)(\tilde \psi_R^\dagger \tilde \psi_L)^2+H.c\nonumber.
\end{align}
The first and last terms respectively favor condensates with $\langle\tilde \psi_R \tilde \psi_L\rangle \neq 0$ and $\langle\tilde \psi_R^\dagger \tilde \psi_L\rangle \neq 0$.  The sign of the latter is arbitrary but, importantly, the former is not: The paired neutral-fermion condensate Josephson couples to the remaining 2D surface and hence its phase is chosen spontaneously but \emph{globally}.  For the 1D subsystem describing the zero modes it thus acts like an external superconductor with a fixed phase, and one should replace $(\tilde \psi_R \tilde \psi_L)^2\rightarrow \tilde \psi_R \tilde \psi_L$.  We then arrive precisely at the Hamiltonian density quoted in Eq.~\eqref{eqn.zeromode}.

\section{Relationship Between Symmetrically Gapped Strong and Weak Topological Insulator Surfaces}\label{stiwti}

Here we comment further on the connections between the symmetric Abelian topological orders for gapped WTI surfaces and the known non-Abelian topological orders for gapped STI surfaces.  For the latter, two phases unrelated by simple condensation transitions are possible \cite{BondersonTO,WangTO,ChenTO,MetlitskiTO}: the T-Pfaffian noted in the main text and the somewhat more elaborate Pfaffian-antisemion phase. Four possibilities for gapping the WTI surface thus immediately arise---each of the two Dirac cone can be gapped independently into either of these phases. As we argue below, we expect that both of the minimal Abelian topological orders that can arise on the WTI surface---corresponding to plates with $K = 4 \sigma^x$ and $4\sigma^z$---may be accessed from these non-Abelian states via condensation transitions.  Which Abelian states arise depends on which of the four possible parent non-Abelian phases one considers.  In particular, for the simplest non-Abelian `double T-Pfaffian' state, obtaining topological order corresponding to $4\sigma^z$ can be ruled out leaving $4 \sigma^x$ as the only natural possibility.

To relate the possible STI and WTI states, suppose first that the WTI surface can be fully gapped by purely \emph{local} interactions between 2D topologically ordered plates and QSH edges (i.e., acting solely within a given interface).  This is precisely the situation we encountered (by construction) in the main text.  In this case the WTI surface inherits only the quasiparticles native to the 2D plates, with no additional topologically distinct excitations generated.  Continuing to assume the minimal two-dimensional K-matrix for the plates, the quasiparticle content of the symmetric topologically ordered WTI surface thus corresponds to either $K = 4\sigma^x$ or $4 \sigma^z$ as discussed below Eq.~\eqref{bigK}.

Suppose next that the symmetrically gapped WTI surface instead descended from a topologically ordered parent state---e.g., the `double T-Pfaffian'---via a simple condensation transition.  From this viewpoint the quasiparticles and their properties are inherited from the parent phase, and simply correspond to the anyons that remain deconfined after the transition.  Again, no additional anyons appear in the process.

We can thus rule out potential parent states by identifying quasiparticles that appear in the descendant state, but not in the putative parent.  Symmetric WTI surface topological orders corresponding to $4 \sigma^x$ and $4 \sigma^z$ both admit quasiparticles with charge $e/2$. In the $4 \sigma^x$ case these excitations carry topological spins $e^{i n \frac{\pi}{2}}$ for integer $n$, while for $4 \sigma^z$ their topological spins are instead given by $e^{i (2n+1) \frac{\pi}{4}}$.
Regarding the parent states, both the T-Pfaffian and Pfaffian-antisemion phases host Abelian $e/2$ excitations with topological spin $\pm i$; the Pfaffian-antisemion (which admits more anyon types) hosts additional $e/2$ quasiparticles with spins $\pm 1$.  Both phases also harbor non-Abelian $e/4$ excitations, which carry spin $1$ in the T-Pfaffian but spin $e^{i \pi/4}$ in the Pfaffian-antisemion \cite{ChenTO}.  Thus in either the `double T-Pfaffian' or `double Pfaffian-antisemion', all quasiparticles with half-integer charge have topological spins of the form $e^{i n \frac{\pi}{2}}$, consistent with $4 \sigma^x$ but not with $4 \sigma^z$. In contrast, a phase where the two Dirac cones are gapped differently---i.e., one in the T-Pfaffian and one in the Pfaffian-antisemion---exhibits $e/2$ excitations with topological spins $e^{i n\pi/4}$. Hence this phase could potentially give way to $4\sigma^z$ upon condensing appropriate combinations of quasiparticles.

We note that the analysis presented in main part of the paper may be interpreted as explicitly carrying out a condensation starting form a symmetric parent state that is closely related to the `double T-Pfaffian'.  In this modified parent state, the two copies of T-Pfaffian topological order are not independently symmetric under time reversal and translations, but instead related by these symmetries; see Fig.~\ref{fig:stack}(b) and Ref.~\cite{STO}. This approach has the technical advantage that it permits a controlled analytic treatment in terms of quasi-1D Hamiltonians and bosonization. Apart from this, we do not expect any significant difference between either method.

\end{document}